\documentclass[fleqn,usenatbib]{mnras}

\usepackage{newtxtext,newtxmath}

\usepackage[T1]{fontenc}
\usepackage{ae,aecompl}

\usepackage{graphicx}	
\usepackage{amsmath}	
\usepackage{amssymb}	

\def\vel{\, \mathbf v}
\def\Bfield{\, \mathbf B}
\def\Efield{\, \mathbf E}

\def\upart{\, \mathbf u_{\rm part}}

\title[3-D NRS instability near astrophysical shocks]{Three-dimensional simulations of non-resonant streaming instability and particle acceleration near non-relativistic astrophysical shocks}

\author[van Marle, Casse \& Marcowith]{
Allard Jan van Marle,$^{1,2}$\thanks{E-mail: vanmarle@apc.univ-paris7.fr}
Fabien Casse,$^{2}$
Alexandre Marcowith,$^{3}$
\\
$^{1}$ Department of Physics, UNIST, UNIST-gil 50, Ulsan, 44919, Korea \\
$^{2}$Universit\'e de Paris, AstroParticle \& Cosmologie,  CNRS \\
CEA, Observatoire de Paris, PSL Research University, CNES, F-75013 Paris, France\\
$^{3}$Laboratoire Univers et Particules de Montpellier (LUPM) Universit{\'e} Montpellier, CNRS/IN2P3, CC72, place Eug{\`e}ne Bataillon,\\ 34095, Montpellier Cedex 5, France.\\
}

\date{Accepted XXX. Received YYY; in original form ZZZ}

\pubyear{2017}

\begin{document}
\label{firstpage}
\pagerange{\pageref{firstpage}--\pageref{lastpage}}
\maketitle

\begin{abstract}
We use particle-in-magnetohydrodynamics-cells to model particle acceleration and magnetic field amplification in a high Mach, parallel shock in three dimensions and compare the result to 2-D models. This allows us to determine whether 2-D simulations can be relied upon to yield accurate results in terms of particle acceleration, magnetic field amplification and the growth rate of instabilities. 
Our simulations show that the behaviour of the gas and the evolution of the instabilities are qualitatively similar for both the 2-D and 3-D models, with only minor quantitative difference that relate primarily to the growth speed of the instabilities. 
The main difference between 2-D and 3-D models can be found in the spectral energy distributions (SEDs) of the non-thermal particles. The 2-D simulations prove to be more efficient, accelerating a larger fraction of the particles and achieving higher velocities. 
We conclude that, while 2-D models are sufficient to investigate the instabilities in the gas, their results have to be treated with some caution when predicting the expected SED of a given shock.
\end{abstract}

\begin{keywords}
plasmas -- methods: numerical -- (magnetohydrodynamics) MHD -- astroparticle physics -- shock waves
\end{keywords}



\section{Introduction}
\label{sec:intro}
Shocks are likely at the heart of the production mechanism of Cosmic Rays (CRs). The investigation of CR acceleration at astrophysical collisionless shocks however requires very demanding numerical resources as the process involve several order of magnitude in scale and time \citep{Marcowith16}. In a first paper, \citet{paper1} hereafter Paper~1, as well as in \citet{vanmarleetal:2017} and \citet{casseetal:2018}, we introduce a new version of the {\tt MPI-AMRVAC} code \citep{vanderHolstetal:2008}, capable of combining grid-based magnetohydrodynamics (MHD) with the particle-in-cell (PIC) method, according to the formalism described by \citet{Baietal:2015} (see also recent contributions by \citet{Mignone18, Amano18}) but including mesh-refinement techniques. We then applied this code to the problem of a magnetized flow passing through a high Mach shock. For the case where the magnetic field is parallel with the direction of the flow we recover the results obtained by \citet{Baietal:2015}, which in turn confirmed earlier models by \citet[e.g.][]{Riquelme10, Caprioli14a} which used a di-hybrid approach (treating the protons as particles but the electrons as a fluid), as well as \citet{Reville08, Reville13}, which used the Vlasov-Fokker-Planck (VFP) method. 
All these simulations show that when supra-thermal particles, accelerated by the shock escape upstream, they interact with the magnetic field and trigger the non-resonant streaming instability (NRS, \citet{Bell04}), which manifests itself in the form of filamentary structures parallel to the flow and the magnetic field lines. However, so far this process has mostly been studied in two dimensions (see the discussion below).

Three-dimensional simulations are important for several aspects of the description of the cosmic ray-fluid system: the dynamics of CR-driven instabilities and the resulting non-linear structures have to be investigated using 3-D simulations in order to capture their geometry \citep{Bell11}; complete studies of CR and magnetic field line transport require a 3-D description of the turbulence \citep{Casse02, Marcowith06, Belletal:2013}, this aspect is especially important to what concerns the way CRs escape from the shock region \citep{Reville12}. Some former 3-D studies have investigated CR-driven streaming instabilities but either with a lower resolution or for shorter timescales or using numerical set-ups without including any shock front. \citet{Lucek00} (based on the techniques developed by \citet{ZacharyCohen:1986}) proposes an analysis of the CR resonant steaming instability using a PIC-MHD method including a calculation of the force exerted by CRs over the fluid solutions. In this study 3-D runs confirm the dynamics of the instability described in 1- and 2-D. However these runs are quite limited in resolution and are designed using a special set up where CR are injected at a box boundary, so without including any dynamical shock structure. \citet{Bell04}, \citet{Zirakashvili08}, \citet{Zirakashvilietal08} investigate the NRS instability using 3-D MHD simulations. The authors investigate the linear growth and the saturation of the instability showing the possibility of strong magnetic field amplification in supernova remnant type shocks. However, in all cases the CR current was kept fixed, and hence no backreaction over the MHD flow was possible to account for \footnote{We also note the work of \citet{Matthews17}, not necessarily connected to a shock configuration which uses 3-D MHD runs including CR currents to investigate the efficiency of NRS modes growth}. \citet{Reville08} first propose a 3-D investigation of the NRS including a detailed calculation of CR transport properties in the magnetized turbulence resulting from the instability. However, CR trajectories are computed in the test-particle limit and the simulations do not include any shock front dynamics. \citet{Bell11} develop a 3-D VFP calculation to investigate shock acceleration and magnetic field amplification for different background magnetic field obliquity. Parallel, oblique and perpendicular shocks all show a destabilization of the NRS instability, but perpendicular shocks are less efficient CR accelerators because the probability of downstream CRs to recross the shock decreases. However also, CR backreaction was not included in this work. \citet{Belletal:2013} and \citet{Reville13} use a modified version of the VFP code coupled with MHD equations through the calculation of the CR current. Using similar set ups they investigate CR transport regimes and expected maximum CR energies for different CR orientation and background magnetic field obliquity. They conclude that in parallel shocks CR diffuse in a sub-Bohm regime (with respect to the background magnetic field) and at a slower rate in oblique shocks. However, once the magnetic field amplification is onset oblique shocks behave like parallel shocks as the magnetic field is highly disordered (see also Paper~1). The VFP technique is powerful but as all techniques designed for large scale shock studies it does not calculate the CR injection from microphysics. \citet{Caprioli14a, Caprioli13} perform hybrid simulations of non-relativistic shocks including a particle-in-cell treatment of ions (thermal and non-thermal). The investigation is conducted mostly in 2-D but also shows some 3-D run analysis. These results confirm an efficient magnetic field amplification concomitant with particle acceleration at parallel shocks. 3-D simulations of strongly oblique (with an angle of $70^{\rm o}$ between the shock normal and the background magnetic field and beyond) do not exhibit any particle acceleration and magnetic field amplification. These simulations are however limited in time and space. The number of macro-particle per cell to calculate both thermal and non-thermal proton dynamics is low (4 particles per cell). In Paper~1 we performed calculations at high-obliquity shocks and found that particle acceleration can proceed in two steps: the first one is associated with a shock-drift process (a result similar to \citet{Caprioli14a} solutions) but once particles have reach some energy threshold a streaming instability is onset leading to diffusive acceleration. It is important to point out that if the PIC-MHD has the advantage of much better statistics to describe the non-thermal population of protons and can probe larger spatial and longer time scales it faces the difficulty to correctly describe non-thermal particle injection only via a parametrization of the energy at which these particles are injected. It is correct to state that PIC-MHD also cannot describe any shock front microphysics and especially the overshoot-undershoot region which particles in highly oblique shocks have to overtake in order to be accelerated diffusively. This important question deserves deep investigations and require proper PIC simulations in highly-oblique shock configuration to be fully elucidated \citep{Ha18}. Work in progress is ongoing to highlight the issue of the injection procedure in PIC-MHD codes. This aspect explains also why this study is only performed in the parallel shock configuration. \\
 
In this work we extend our simulations to 3-D and investigate if this changes 2-D the results obtained using the PI[MHD]C code developed in Paper~1 in parallel shock configuration. In particular, we wish to determine whether the filamentary structure observed in 2-D models of the NRS will translate into a tube-like structure in a 3-D simulation. We also produce long simulations with the aim to capture long-term behaviors of CR modified shocks. In particular we add a shock-capturing procedure to extend particle injection process in strongly corrugated shocks. \\

In section \ref{S:NUM} we recall the main principles of coupling PIC and MHD techniques. We also describe the shock capturing module used for long term particle injection. In particular in section \ref{S:SIM} we present the different numerical set ups used in this study. Results are discussed in section \ref{S:RES} before concluding in section \ref{S:CONC}.

\begin{figure*}
\centering
\mbox{
\includegraphics[width=0.45\textwidth]{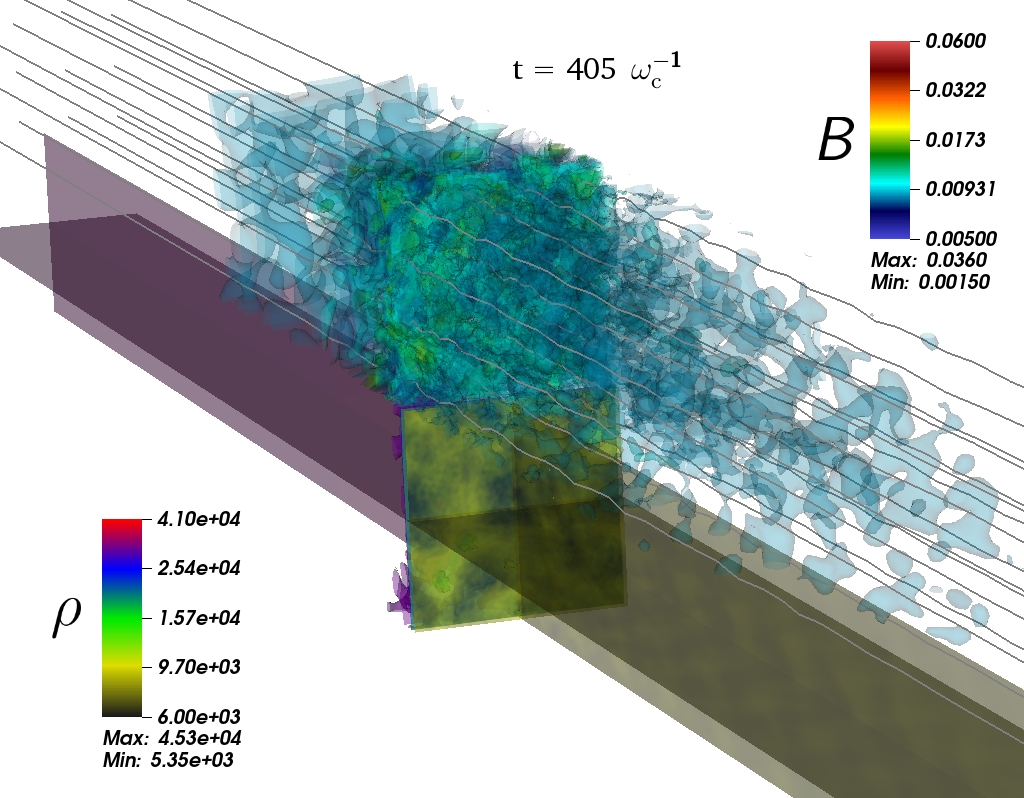}
\includegraphics[width=0.45\textwidth]{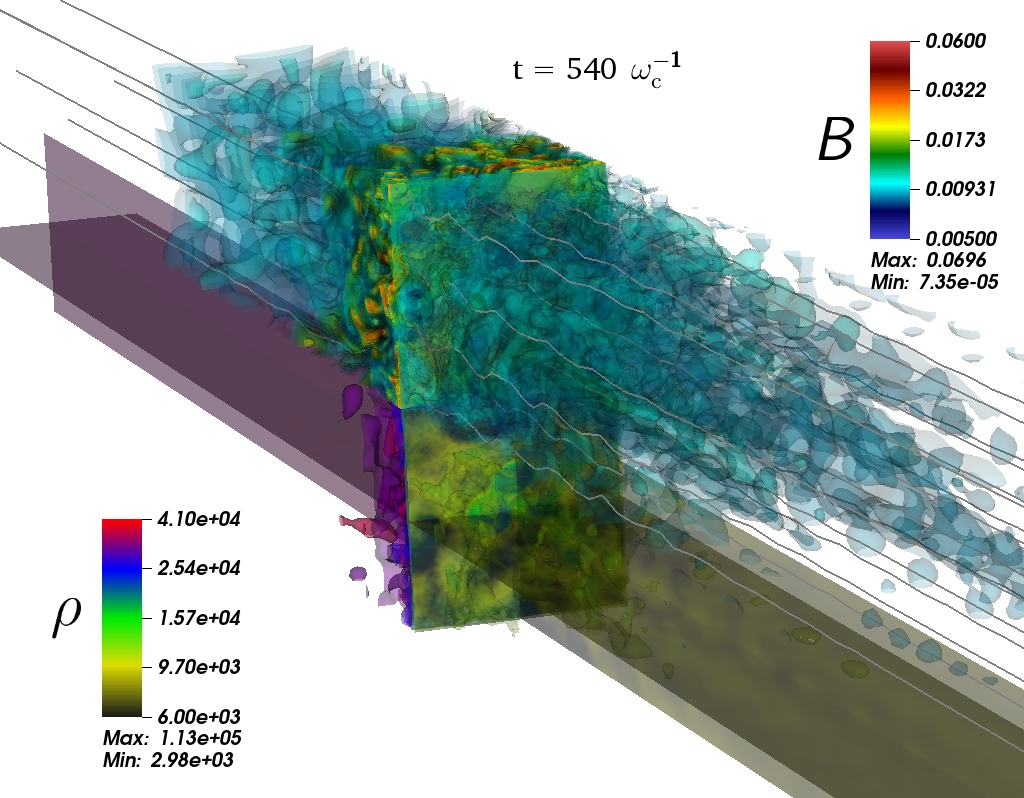}} \\ 
\mbox{
\includegraphics[width=0.45\textwidth]{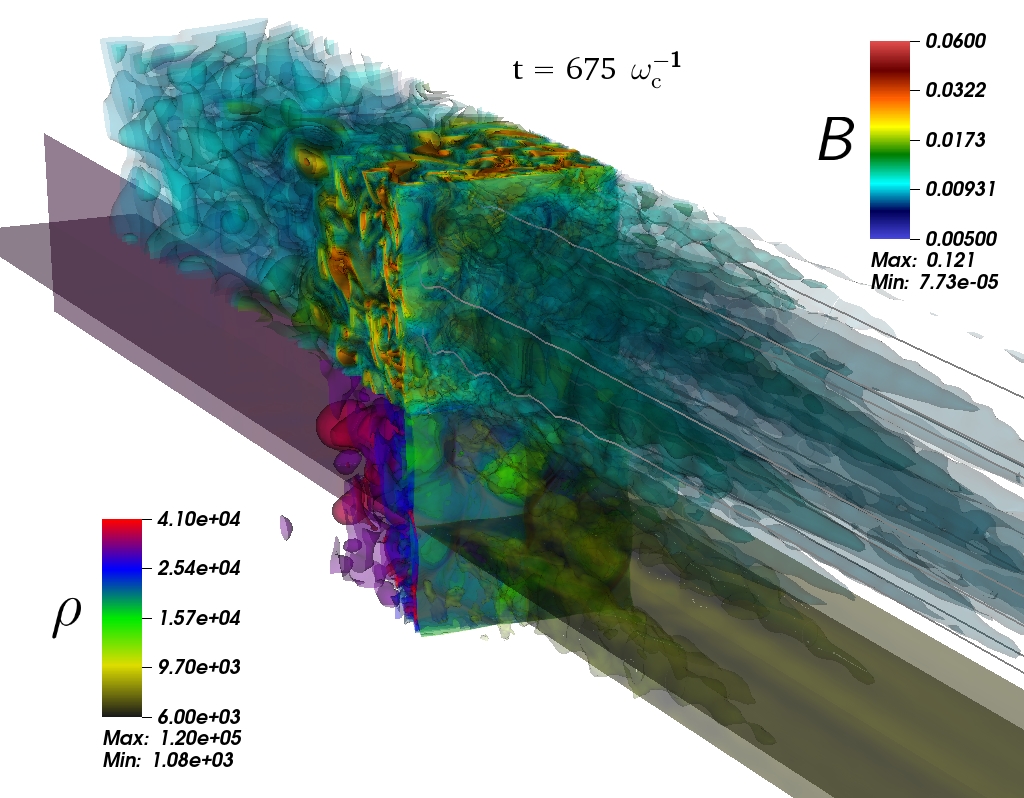}
\includegraphics[width=0.45\textwidth]{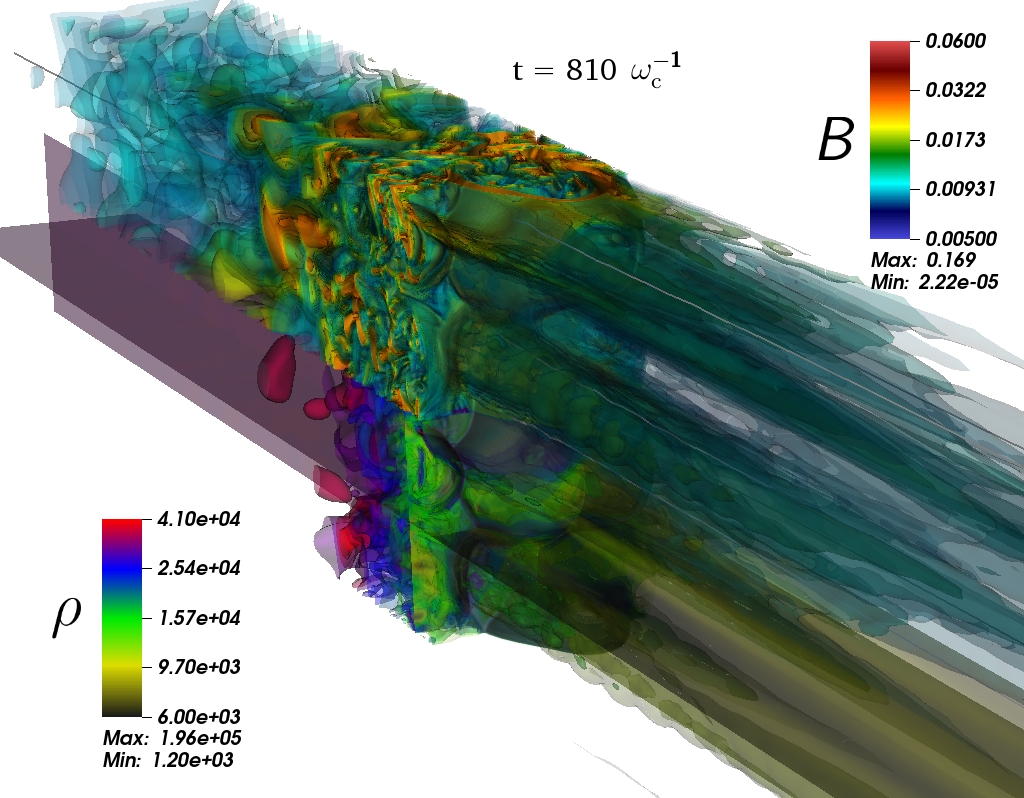}}
\caption{Magnetic field strength and thermal plasma density at four consecutive moments in time, showing the development of a filamentary instability in the upstream medium as well as the distortion of the shock front. Initially, (t=405 $\omega_{\rm c}^{-1}$, upper left panel) the structure in the magnetic field seems random and the thermal gas density remains undisturbed. However, over time (t=540, 675 and 810 $\omega_{\rm c}^{-1}$) the disturbance in the upstream magnetic field forms filaments and the thermal gas density follows. The downstream magnetic field becomes highly turbulent. Owing to the varying pressure, the shock front becomes distorted}
 \label{fig:3Devolution}
\end{figure*}

\section{Numerical method and set-ups}
\label{S:NUM}
\subsection{General statements}
The combined PIC-MHD method \citep{Baietal:2015, Mignone18, Amano18} and Paper~1, from here on referred to as particles in MHD cells (PI[MHD]C), is based on the assumption that the plasma can be described as a thermal plasma, in which a small fraction of the particles behave in a non-thermal fashion. 
Therefore, it can be treated numerically by using traditional, grid-based MHD to describe the thermal plasma and the magnetic field, whereas the non-thermal component is calculated using the PIC method. 
The interaction between the two components is taken into account through a modified version of Ohm's law,
\begin{equation}\label{Eq:ELE}
c\Efield = -\left((1-R)\vel +R\upart\right)\times\Bfield
\end{equation}
with $c$, the speed of light, $\Efield$ the electric field, $\vel$ the velocity of the thermal plasma, $\upart$ the average velocity of the supra-thermal particles, $\Bfield$ the magnetic field and $R$ the ratio of the supra-thermal particle charge density to the total charge density ($R\ll 1$). MHD momentum and energy equations also consider contributions from the non-thermal population leading to a global momentum and energy conservation. The gas is considered to be non-collisional and the only interaction between the non-thermal particles and the thermal gas is through the electro-magnetic field. \\

We use the same code described in Paper~1, which is based on the {\tt MPI-AMRVAC} code \citep{vanderHolstetal:2008,Keppensetal:2012}. This is an MPI-parallel, fully conservative code that uses the {\tt OCTREE} \citep{ShephardGeorges:1991} adaptive mesh system to dynamically adapt the grid resolution. 
Onto this code we have added a module that calculates the motion of charged particles in an electro-magnetic field using the Boris-pusher \citep{BirdsallLangdon:1991}. 
The influence of the charged particles on the thermal gas is accounted for through the modified conservation equations described by \citet{Baietal:2015, Mignone18, Amano18} and Paper~1. 
This method allows for a self-consistent interaction between the particles and the thermal gas while conserving momentum and energy. 
In order to ensure that the magnetic field remains divergence-free throughout the simulation we have implemented the constrained-transport method as described in \citet{Balsara:1998,BalsaraSpicer:1999}.\\

\subsection{Corrugated shock capture procedure}
In the simulations shown in Paper~1 we injected the particle along a straight line, perpendicular to the flow. The position of this line was determined by the x-coordinate (hereafter the coordinate along the shock normal) of the highest density gradient, which was assumed to coincide with the location of the shock. However, once the shock becomes corrugated, this assumption is no longer valid. Therefore we have now opted for a different approach.
In order to optimize the injection process, we determine the position the shock at the beginning of each time step. We do this by calculating the sonic Mach number(in the background frame of reference) in each grid point to determine where the plasma makes a transition from supersonic to subsonic. Together these grid points form an iso-surface that is saved at each time step. The particles are injected randomly inside a volume determined by this iso-surface and a distance of one Larmor radius in the post-shock direction. 
This allows us to ensure that each particle is injected within a distance of one Larmor radius downstream of the shock, even when the shock becomes warped as a result of upstream instabilities and downstream turbulence.

\begin{figure*}
\centering
\mbox{
\includegraphics[width=0.45\textwidth]{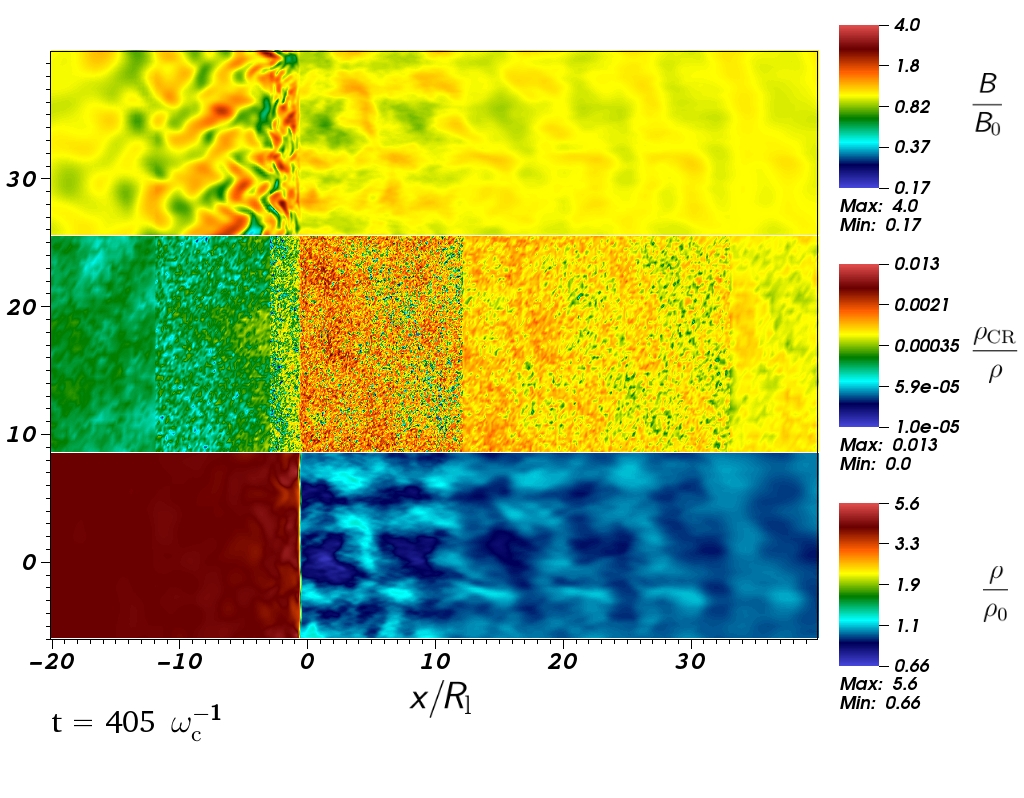}
\includegraphics[width=0.45\textwidth]{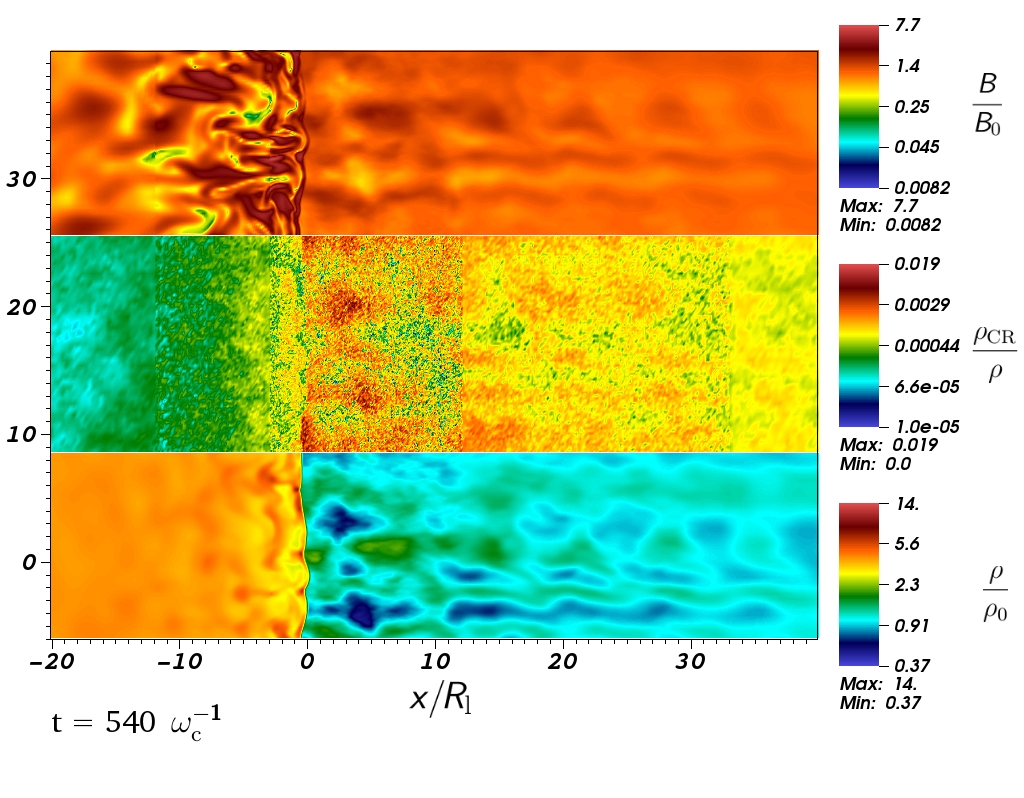}} \\ 
\mbox{
\includegraphics[width=0.45\textwidth]{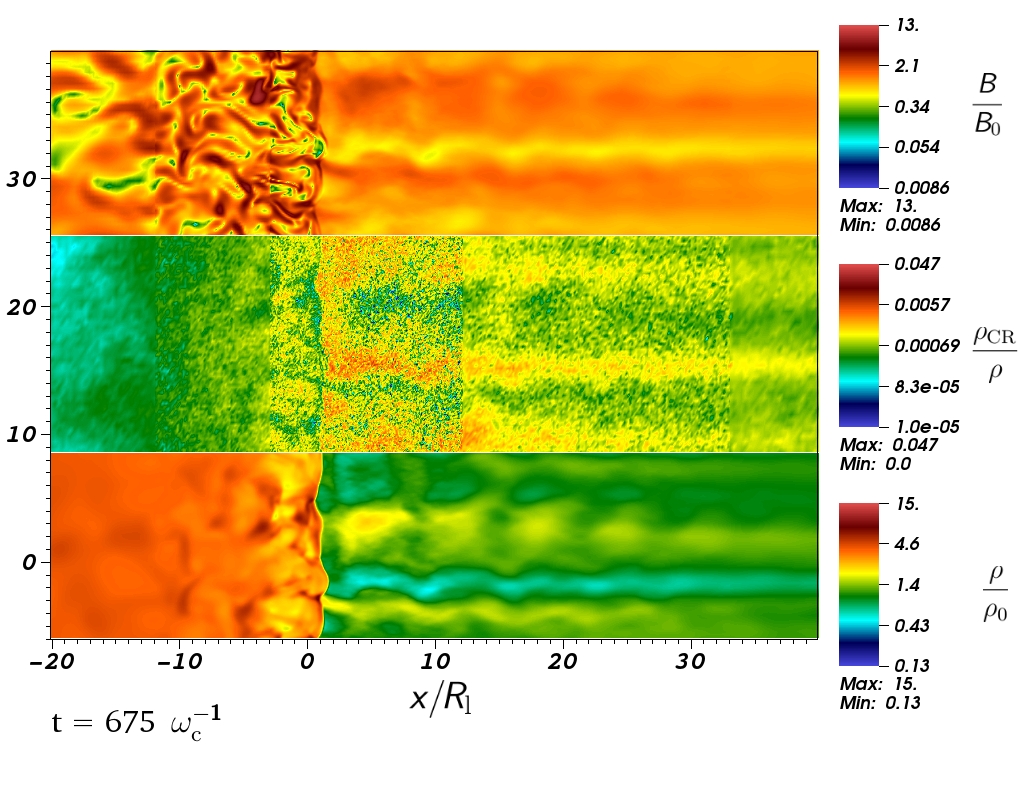}
\includegraphics[width=0.45\textwidth]{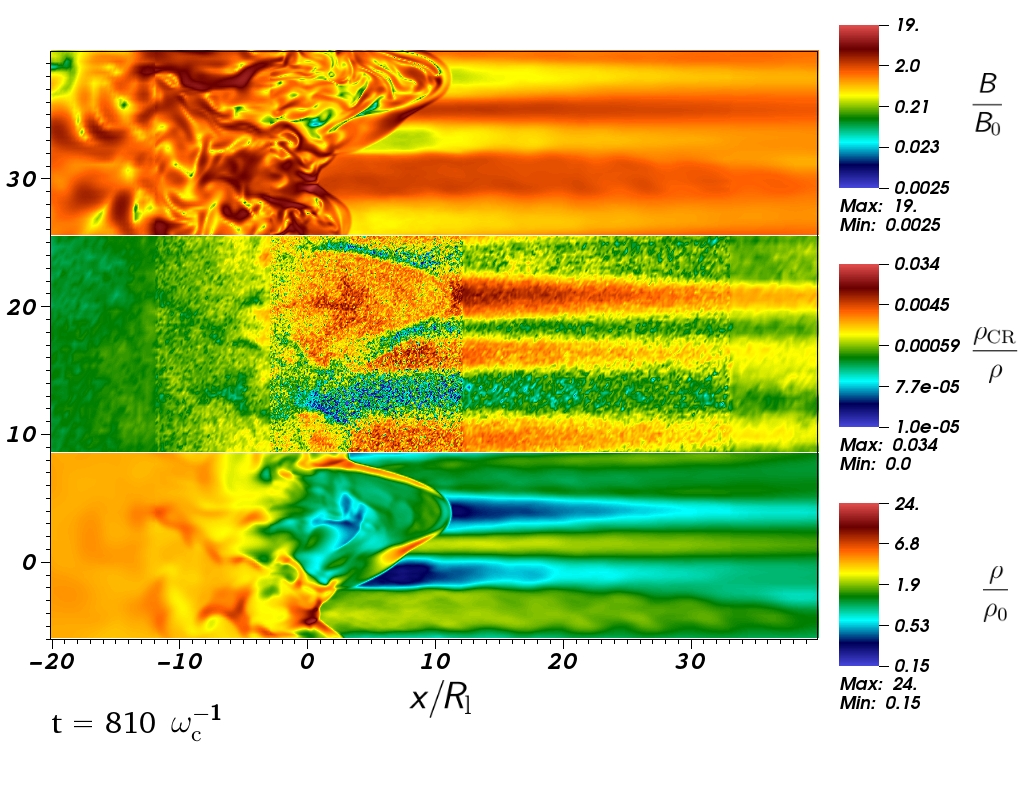}}
\caption{ Magnetic field strength relative to the initial magnetic field strength $B_0$ (top), relative  density of the non-thermal particles compared to the thermal gas (middle), and thermal gas density relative to the initial upstream thermal gas density $\rho_0$ (bottom) at same moments in time as in Fig.~\ref{fig:3Devolution}. Each figure shows a cut through the simulation box along the yz-diagonal. Note that his image shows a zoom-in on the shock. The actual simulation box extends along the x-axis from $-96$ to $96 R_l$.}
 \label{fig:3Devolutioncut}
\end{figure*}

\subsection{Three-dimensional simulations set-up}
\label{S:SIM}
We repeat the simulation of a parallel Alfv\`enic Mach number $M_{\rm A}$\,=\,30 shock as presented in Paper~1 where the velocity of the shock is set to $3\times 10^{-3}c$ ($c$ being the velocity of light). 
In this case, we use a 3-D box measuring $192\times\,12\,\times\,12\,R_{\rm l}$, with $R_{\rm l}$ the Larmor radius defined by the magnetic field and the injection velocity of the particles. As in Paper~1 we start the simulation from the analytical solution for a standing shock as prescribed by the Rankine-Hugoniot conditions with the gas moving along the x-axis and the shock being at the centre of the grid. We assume that the upstream gas is in equipartition (magnetic energy density equals thermal energy density). From this initial condition we start to introduce the particles (ions only, we assume all electrons are thermal) at the shock front, at a rate that conforms to 0.2 percent of the mass passing through the shock (as suggested by PIC simulations, 
see e.g. \citet{Caprioli14a} ). 
The particles are introduced at a velocity $v_{\rm inj}$ equal to 3 times the pre-shock velocity and are moving isotropically relative to the post-shock fluid. The simulation runs for a period of $t\,=\,10^3 R_{\rm l}/v_{\rm inj}$ where $R_{\rm l}/v_{\rm inj}$ is the inverse of the ion cyclotron pulsation $\omega_{\rm c}^{-1}$ related to the ion plasma pulsation $\omega_{\rm pi}= \omega_{\rm c}c/V_A$ ($V_A$ being the Alfv\`en speed). Over this period we inject a total of $10^8$ particles.

The grid has a basic resolution of $640\times40\times40$ grid cells. In the area directly around the shock (one Larmor radius downstream and 10 Larmor radii upstream) we allow two levels of refinement, creating an effective resolution of 13.3 grid cells per Larmor radius (equivalent to a local resolution of $2560\times160\times160$ grid cells). Further from the shock (up to 10 Larmor radii downstream and 30 Larmor radii upstream) we allow one level of refinement, for a resolution of 6.7 Larmor radii per grid cell. Unlike our previous work, we fix the grid at the start of the simulation and do not allow it to adapt as the simulation progresses. Such choice stems from the fact that enforcing the highest resolution far from the shock does not significantly impact the outcome of the simulations while significantly increasing the numerical cost.

We set the upstream MHD boundary to maintain a steady inflow of gas and allow a free outflow at the downstream boundary so that the gas flows along the x-axis in the negative direction. The y- and z-boundaries are considered periodic. For particles, the y- and z- boundaries are also periodic. However, both x-boundaries allow the particles to escape from the simulation box. This contrasts with PIC and PIC-hybrid simulations, which typically involve creating a shock by allowing a beam of particles to collided with a solid wall that is reflective to both thermal and non-thermal particles. 

In order to check our results we perform two simulations in 2-D with an identical physical box size, spatial resolution, and input parameters, but with a different particle weight. For the first simulation we inject a total number of 2.5$\times10^{7}$ particles over a period of 1000 $\omega_{\rm c}^{-1}$. For the second, we inject 5$\times10^{7}$ over the same period of time. In both cases the total mass fraction that is injected remains the same.

\begin{figure*}
\includegraphics[width=\textwidth]{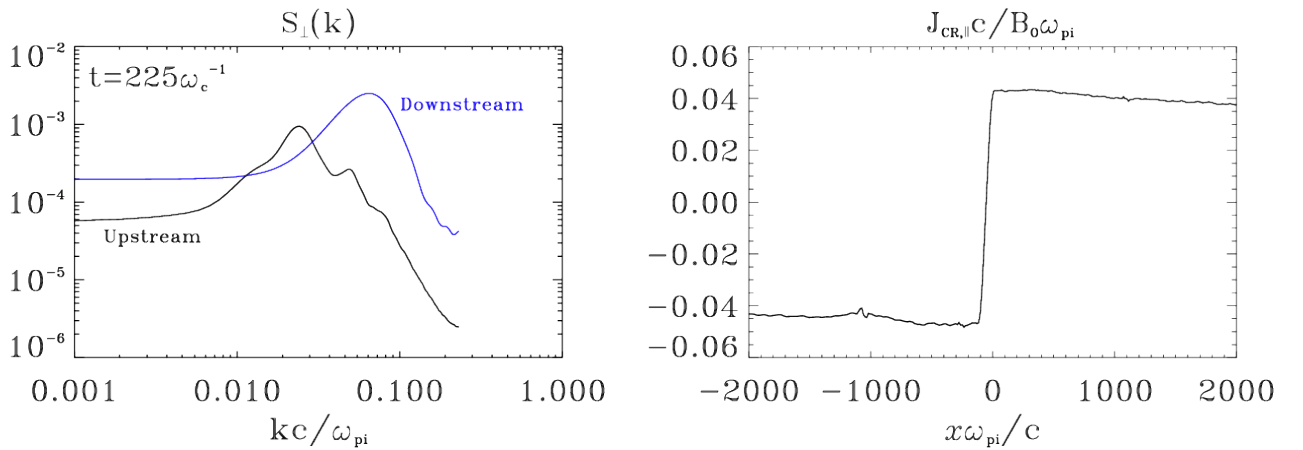}
\caption{{\bf Left:} Magnetic power spectrum in both the upstream and downstream media as a function of the parallel wave vector $k$. {\bf Right:} Parallel cosmic-ray current as a function of the distance to the front shock. Here $x\omega_{\rm pi}/c\,=xv_{\rm inj}/R_lV_A$ with $v_{\rm inj}=3\times M_AV_A$, so the figure covers the area from $x\,=\,-22.2R_l$ to $+22.2 R_l$. The upstream spectrum peaks at $k\sim 0.02 \omega_{\rm pi}/c$ which is in agreement with the expected maximal wave vector growth value from the non-resonant streaming instability, namely $k_{\rm max}=J_{CR,\parallel}/2B_0$.}
\label{fig:spectrum}
\end{figure*}

\begin{figure}
\includegraphics[width=\columnwidth]{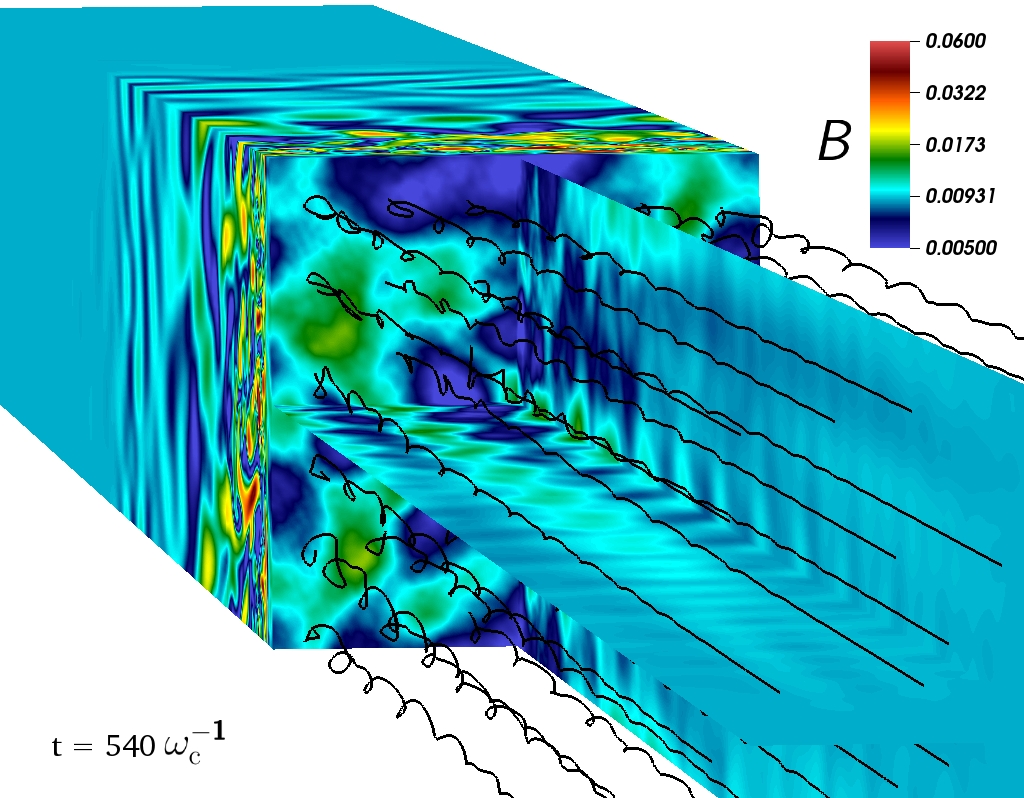}
\caption{Absolute magnetic field strength and upstream magnetic field lines at $t=540\, \omega_{\rm c}^{-1}$, showing the loops in the upstream magnetic field caused by the streaming instability.}
\label{fig:Bfield}
\end{figure}

\section{Results}\label{S:RES}
\subsection{3-D simulations of NRS instability near parallel shocks}
As was already briefly discussed in \citet{vanmarleetal:2017} the 3-D simulation shows the same trend that was observed in the 2-D models in Paper~1 as well as earlier models by \citet[e.g.][]{Caprioli14a,Baietal:2015}. Upstream, the supra-thermal particles trigger the NRS, which manifests itself in the form of filamentary structures parallel to the streamlines (see Fig.~\ref{fig:3Devolution}-\ref{fig:3Devolutioncut}) similar to the structures observed by \citet{Baietal:2015}.
These filaments are initialized near the shock and extend into the upstream medium, increasing in length over time.
Initially, this shows primarily in the magnetic field (Fig.~\ref{fig:3Devolution}, upper panels), but over time the thermal plasma density follows the same pattern (Fig.~\ref{fig:3Devolution}, lower panels). Whereas such structures in a 2-D model are effectively slabs, in 3-D it is shown that they form tubular filaments. Downstream, where the flow is subsonic (if still super-Alfv{\'e}nic) the medium becomes more randomly turbulent. As the upstream medium loses its homogeneity, the ram pressure felt by the shock starts to vary, both in space and time. This leads to a distortion of the shock surface after at time t$=500-600\, \omega_{\rm c}^{-1}$ (Fig.~\ref{fig:3Devolution}, lower panels). Figure~\ref{fig:3Devolutioncut} shows the magnetic field strength relative to the unperturbed magnetic field ($B_0$), the supra-thermal particle density relative to the thermal particle density and the  thermal gas density density relative to the initial preshock thermal gas density ($\rho_0$) for cuts through the 3-D simulation box along the yz-diagonal, in the area near the shock, at the same moments in time as Fig.~\ref{fig:3Devolution} which allows for a closer investigation of the instabilities and the time evolution of the shock as well as comparison to 2-D results (see next section). 

\begin{figure*}
\centering
\includegraphics[width=0.8\textwidth]{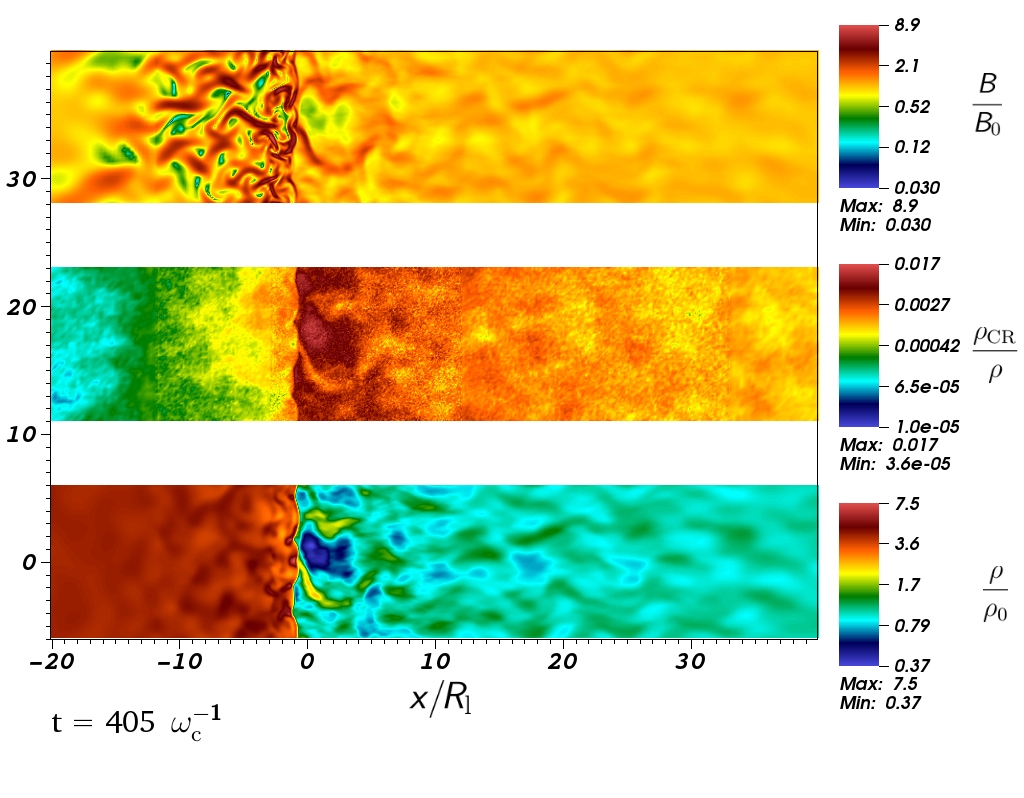}\\
\includegraphics[width=0.8\textwidth]{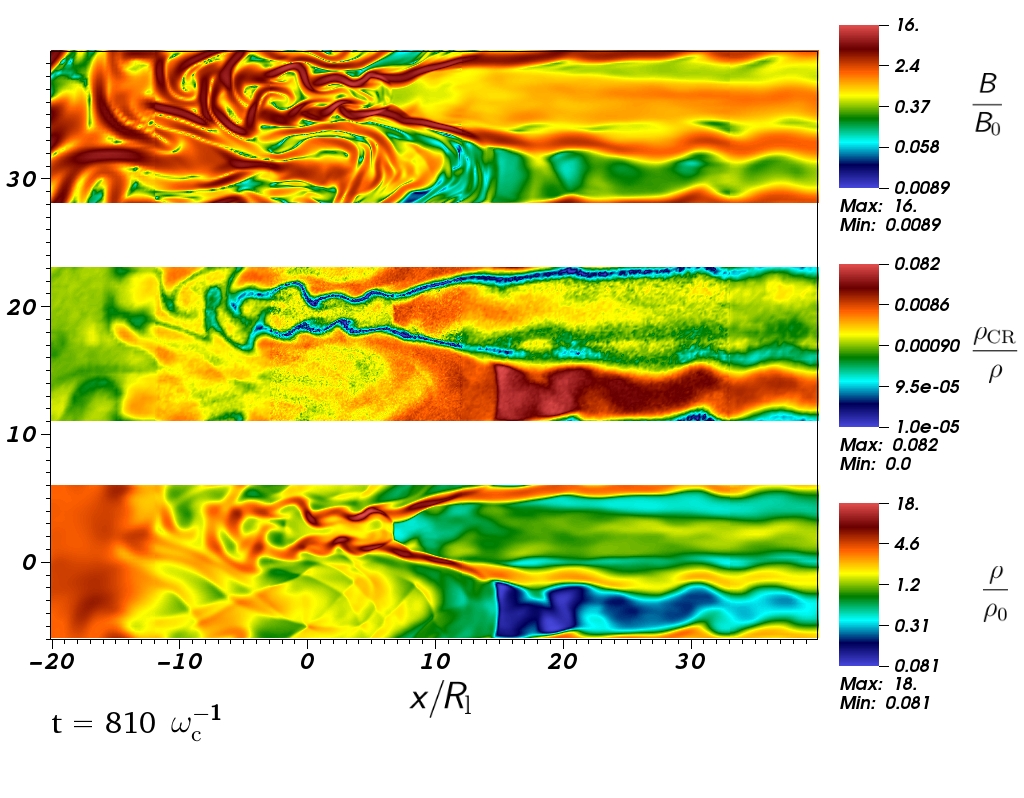}
\caption{For comparison: Snapshots of the 2-D simulations with 2.5 million particles (top) and 5 million particles (bottom), at $t\,=\,405$ and $t\,=\,810\,\omega_{\rm c}^{-1}$ using the same variables as in Fig.~\ref{fig:3Devolutioncut}. The 2D simulation shows a similar morphology as the cuts through the 3D simulation.}
 \label{fig:2DevolutionA}
\end{figure*}

\begin{figure*}
\centering
\includegraphics[width=0.8\textwidth]{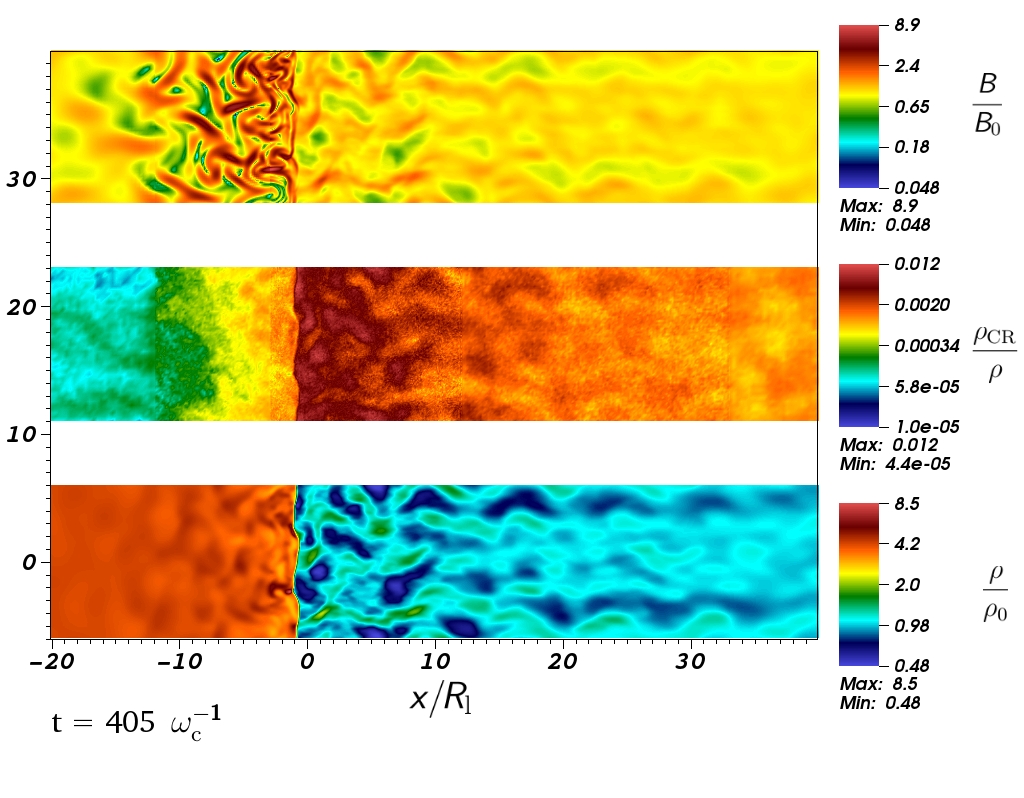}\\
\includegraphics[width=0.8\textwidth]{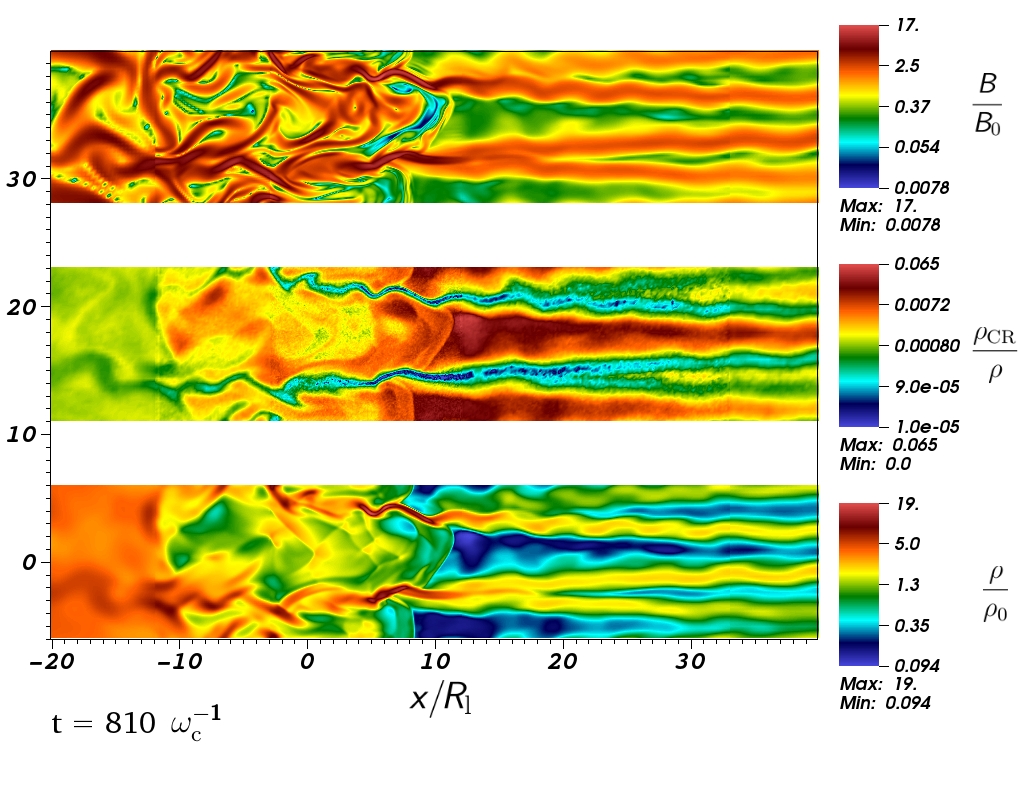}
\caption{For comparison: Snapshots of the 2-D simulations with 5 million particles (bottom), at $t\,=\,405$ and $t\,=\,810\,\omega_{\rm c}^{-1}$ using the same variables as in Figs.~\ref{fig:3Devolutioncut} and \ref{fig:2DevolutionA}.}
 \label{fig:2DevolutionB}
\end{figure*}

Eventually, as the filaments become more pronounced, the ram-pressure experienced by the shock starts to vary and the shock begins to corrugate (bottom right panels in Figs~\ref{fig:3Devolution}-\ref{fig:3Devolutioncut}). However, when observing this we should keep in mind that the injection rate of non-thermal particles is constant in our simulation. It has been shown \citep{Caprioli14a} that the injection rate depends on both the Mach number of the shock and the angle between the shock and the magnetic field. As the shock corrugates, the effective Mach number deviates from the original solution. Furthermore, the angle between the magnetic field and the shock is no longer constant. How this would influence the injection rate is a complicated issue that goes beyond the scope of this paper. Therefore we stop the simulation at this point. However, this run represents to our knowledge the longest simulation describing a 3-D CR-mediated shock structure including both up- and downstream media. 

The validity of our results can be confirmed by comparing the wavelengths of the instabilities in the early phase against analytical predictions \citep{Bell04}. We do this by taking a snapshot of the grid and performing a 2-D Fourier analysis on yz-cuts through the simulation box upstream and downstream of the shock. The result is shown in Fig.~\ref{fig:spectrum}. Upstream, the spectrum peaks at a wavenumber $k\sim 0.02\, \omega_{\rm pi}/c$. This is in agreement with the expected maximal wave vector growth value from the non-resonant streaming instability, namely $k_{\rm max}=J_{CR,\parallel}/2B_0$ \citep{Bell04}, confirming the validity of the model. Here $J_{CR,\parallel}$ and $B_0$ represent the CR current parallel to the shock normal and the strength of the background magnetic field. As in paper I we notice that downstream the scale of the main growing mode is compressed by a factor $\sim$4 as expected near a strong non-relativistic shock. 

The nature of the instabilities can be observed in Fig.~\ref{fig:Bfield}, which shows the upstream magnetic field lines at $t=540\, \omega_{\rm c}^{-1}$. this demonstrates the spiral loops in the upstream magnetic fields, which are characteristic of the streaming instability.

\begin{figure}
\includegraphics[width=\columnwidth]{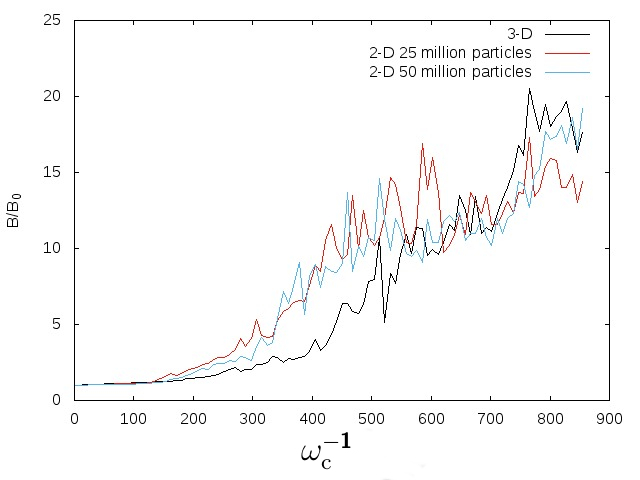}
\caption{Maximum value of the magnetic field amplification as a function of time for both the 3-D model and the 2-D models with $2.5\,\times\,10^7$  and $5.0\,\times\,10^7$particles. the 2-D models shows an earlier increase, but also level off early, whereas the 3-D model shows less initial growth, but eventually starts to surpass the 2-D models.}
\label{fig:bmax}
\end{figure}
\begin{figure*}
\centering
\mbox{
\includegraphics[width=0.33\textwidth]{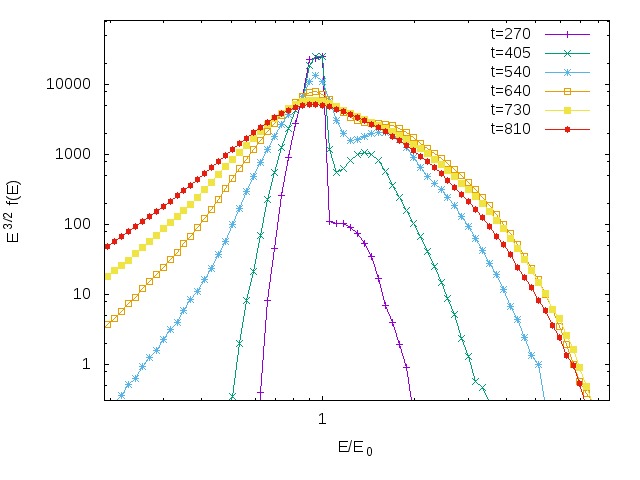}
\includegraphics[width=0.33\textwidth]{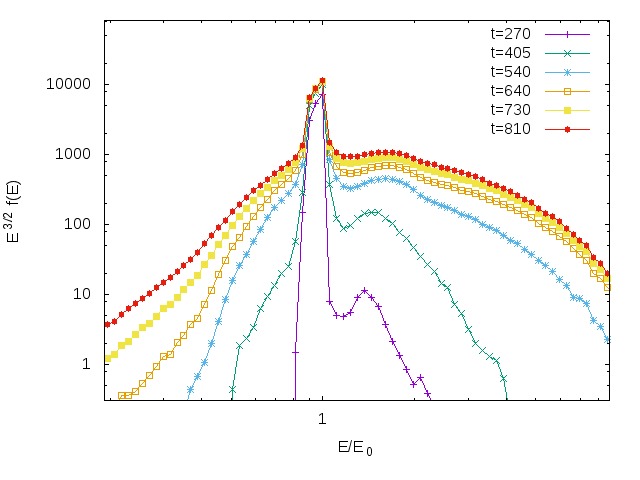}
\includegraphics[width=0.33\textwidth]{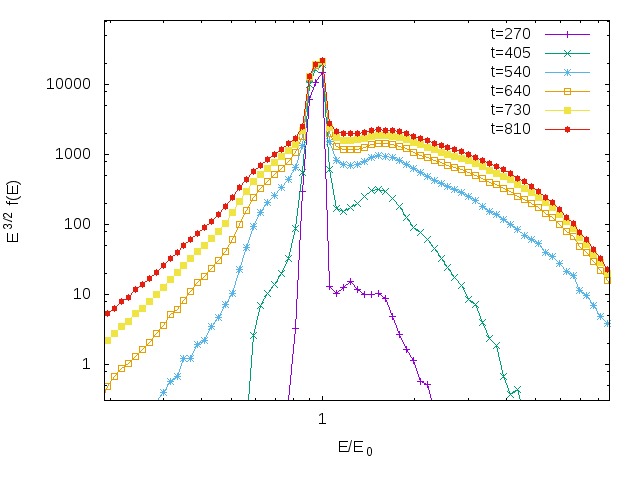}
}
\caption{Time evolution of the non-thermal particle SEDs for the 3-D simulation (left)  25 million particle 2D simulation (centre), 50 million particle 2-D simulation (right). $E_0$ is the particle injection energy.}
\label{fig:seds}
\end{figure*}


\subsection{Comparisons between 2-D and 3-D runs}
 In order to compare the 3-D results with 2-D models, we repeat the simulation in 2-D  as shown in in Figs~\ref{fig:2DevolutionA}-\ref{fig:2DevolutionB}, using the same input parameters and resolution. These simulations are effectively the same as those shown in Paper~1, albeit with a fixed grid and a smaller simulation box. The results are shown in Fig.~\ref{fig:2DevolutionA}, showing the thermal gas density, magnetic field strength and relative non-thermal particle density at $t=405\, \omega_{\rm c}^{-1}$ (left panels) and $t=810\, \omega_{\rm c}^{-1}$ (right panels) for the simulations with $2.5\,\times\,10^7$ particles over a period of $t=1000\, \omega_{\rm c}^{-1}$ and Fig.~\ref{fig:2DevolutionB}, which is identical but for a simulation with $5\,\times\,10^7$ particles over the same time period. The results match qualitatively with the 3-D simulation, showing the same upstream instabilities and shock distortion.\\

Figure~\ref{fig:bmax} shows the evolution of the magnetic field amplification as a function of time for the 3-D simulation compared to the 2-D simulations. For the 2-D simulation, the magnetic field amplification rises earlier, but levels off after t$\simeq 600\, \omega_{\rm c}^{-1}$, whereas for the 3-D model, the amplification remains smaller initially, but then grows to equal and even surpasses that of the 2-D models at the end of the simulation. The latter rise has some impact on particle acceleration discussed in the next section. 

Compared to the 2-D results in \cite{vanmarleetal:2018}, which used the same physical input parameters, the filaments show less structure. This is a consequence of the lower resolution in the areas far from the shock, which was required to make the 3-D model feasible.

\subsection{Particle energy distribution}
\label{sec-sed}
In Fig.~\ref{fig:seds} we show the spectral energy distribution (SED) as function of the normalized CR energy $E/E_0$ in the downstream region. $E_0=m_pv_{\rm inj}^2/2$ stems for the injection energy ($m_p$ being the proton mass). For a high-Mach, non-relativistic shock, the particle-SED is expected to conform to a power-law with a slope that matches $f(E)\propto E^{-3/2}$ \citep{Caprioli14a} and, initially, the particle SED in our 3-D simulation (Fig.~\ref{fig:seds}, left panel) evolves toward that slope. However, after t$\simeq 650\,\omega_{\rm c}^{-1}$, the shock becomes distorted and the magnetic field gets strongly amplified. This reduces the efficiency of the particle acceleration process and causes a deviation from the analytical prediction because the assumptions on which this model is based (a Mach number that is constant in space and time, as well as a fixed angle between the shock and the magnetic field) are no longer valid. 

Not only does the change in shock strength and obliquity change the injection rate, the shock distortion also influences the acceleration process. The distortion of the shock increases its surface area. This increases the chance of a particle crossing the shock surface, but also causes the newly injected particles to be spread out over a larger volume, reducing the effective non-thermal particle density, which in turn influences the local instabilities. Furthermore, the distorted shape of the shock decreases the effectiveness of the acceleration when a particle crosses the shock. The energy a particle gains from each transition across the shock surface depends on the direction of its velocity vector compared to the velocity vector of the flow. In order to gain the maximum momentum, the particle has to move parallel to the flow when crossing from the upstream to the downstream region and anti-parallel to the flow while moving downstream to upstream. 
As long as the shock surface is smooth and perpendicular to the flow this is exactly what happens and the acceleration process is efficient. 
However, once the shock becomes sufficiently distorted, particles will be able to cross the shock surface while moving quasi-perpendicular to the flow or, in extreme cases,  while moving counter to the preferred direction, in which case they will actually lose momentum with each reflection.
The situation is made even more complicated by the fact that the turbulent motion in the downstream medium causes great variation in the local velocity field, which in turn changes the forces acting on the particles.

This is reflected in the behaviour of the SED, which shows a steeper slope than predicted by the analytical model. This effect is far less pronounced in the case of the 2-D simulations. Initially, the SEDs of the 2-D simulations show a similar behaviour that closely matches the 3-D model, but after $T\simeq400\,\omega_{\rm c}^{-1}$ they continue to accelerate particles more efficiently, which is reflected in a more extended high-energy tail. The SEDs of the two 2-D simulations resemble each other closely, showing that the difference in the number of particles does not lead to any significant difference in the end result.

\begin{figure}
\centering
\mbox{
\includegraphics[width=0.95\columnwidth]{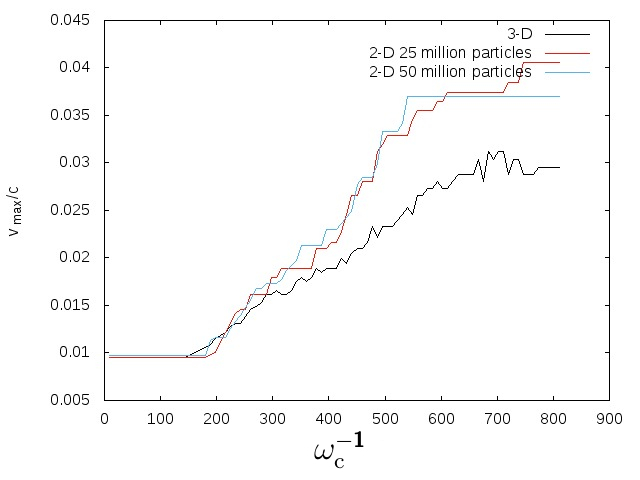}}
\caption{Maximum velocity of the particles in the grid as a function of time relative to the injection energy. The energy of the particles in the 3-D simulation increases more slowly and tends to level off more quickly compared to the 2-D models, because the 3-D simulation is more sensitive to shock distortion.}
 \label{fig:vmax}
\end{figure}

This trend is also reflected in the maximum acceleration that the simulations achieve. Figure \ref{fig:vmax} shows the evolution of the maximum velocity of the particles in the grid. This shows a clear difference between the 2-D and 3-D models, with the 2-D simulations accelerating particles to higher velocity. This is caused by the limitations of the 2-D model, which allows particles to become ``trapped'' in the loops of the magnetic field, where they can be repeatedly accelerated, whereas in 3-D such particles have an easier time escaping from the shock region. 
Although this does create an artefact, it is largely irrelevant to the evolution of the gas because it only happens to a relatively small fraction of the particles (N.B. The SEDs in Fig.~\ref{fig:seds} are all shown in a log-log plot), but should be kept in mind when evaluating SEDs produced by simulations. 

The fact that we simulate in the rest-frame of the shock adds to the difficulty of forming an extended Fermi-plateau in the particle SED because particles are allowed to escape from the simulation in both the downstream and upstream region. 
Simulations that are run in the rest-frame of the downstream region use a reflective wall, which keeps the particles in the simulation box. Since faster particles are more likely to reach the upper and lower x-boundary in our model, they are preferentially removed from the simulation, reducing the extent of the Fermi-plateau. 
The escape of particles at the boundaries changes the current because the particles are no longer being reflected toward the shock. However, The escape of particles at the boundaries changes the current because the particles are no longer being reflected toward the shock. However, this does not seem to influence the result as the morphology of the thermal plasma in our simulations is qualitatively similar to the results of \cite{Baietal:2015}, which were obtained with a reflective boundary.

\section{Discussion and conclusion}\label{S:CONC}
Our results show that the 3-D models follow the same trend as the earlier 2-D simulations. The upstream medium forms the filamentary structures predicted by the analytical model \citep{Bell04}, whereas the downstream medium becomes turbulent. As a result, the shock front becomes corrugated, conforming to the spatial variations in the ram-pressure exerted by the upstream gas. In both 2-D and 3-D models, the distortion changes the nature of the shock, causing local variations in the sonic and magnetic Mach number and varying the angle between the shock surface and the magnetic field. This will influence the injection rate of non-thermal particles near the shock surface. This confirms the 2-D results and indicates that 2-D simulations can be used reliably to model the interaction between non-thermal particles and a collisionless shock. 


Although the behaviour of the thermal gas and magnetic field, as well as the evolution of the instabilities in 2-D and 3-D models is qualitatively the same and even matches quantitatively to a reasonable degree, the particle energy distributions are different, with the 2-D simulations showing a harder spectrum that reaches higher energies. This effect is largely due to the 3-D geometry which includes a supplementary degree of freedom in particle's motion. The 3-D geometry also creates a strong magnetic field amplification, which, at the end of the simulation, surpasses the magnetic field strength found in the 2-D models.\\

The 3-D simulations presented in this study have been conducted up to $1000\,\omega_{\rm c}^{-1}$ and then remain the longest to date describing a CR-mediated collisionless shock. They confirm the results obtained by previous hybrid simulations that in these shocks particle acceleration is mediated by the onset of the NRS instability. \\

Even if we have improved the shock capturing with respect to paper I a main limitation for a long term description of the CR-magnetized fluid system remains due to a lack of a proper injection procedure accounting for the strong shock distortion. A detailed investigation of the injection recipe to be adopted is necessary to a better evaluation of the SED and the evolution of the maximum energy in 3-D runs. An aspect to be treated in a forthcoming work where different injection recipes including the back-reaction of the accelerated CRs will be tested.

\section*{Acknowledgements}
This work acknowledges financial support from the UnivEarthS Labex program at Universit\'e de Paris  (ANR-10-LABX-0023 and ANR-11-IDEX-0005-02) and from the NRF of Korea through grant 2016R1A5A1013277 and by the Basic Science Research Program through the National Research Foundation of Korea (NRF) funded by the Ministry of Education, Science and Technology (2018R1D1A1B07044060).
This work was granted access to HPC resources of CINES under the allocation A0020410126 made by GENCI (Grand Equipement National de Calcul Intensif). 
We are grateful to our anonymous reviewer for the helpful comments that assisted us in improving our paper.




\bibliographystyle{yahapj}
\bibliography{vanmarle_biblio.bib}

\appendix
\section{Animations}

\begin{figure}
\centering
\mbox{
\includegraphics[width=0.95\columnwidth]{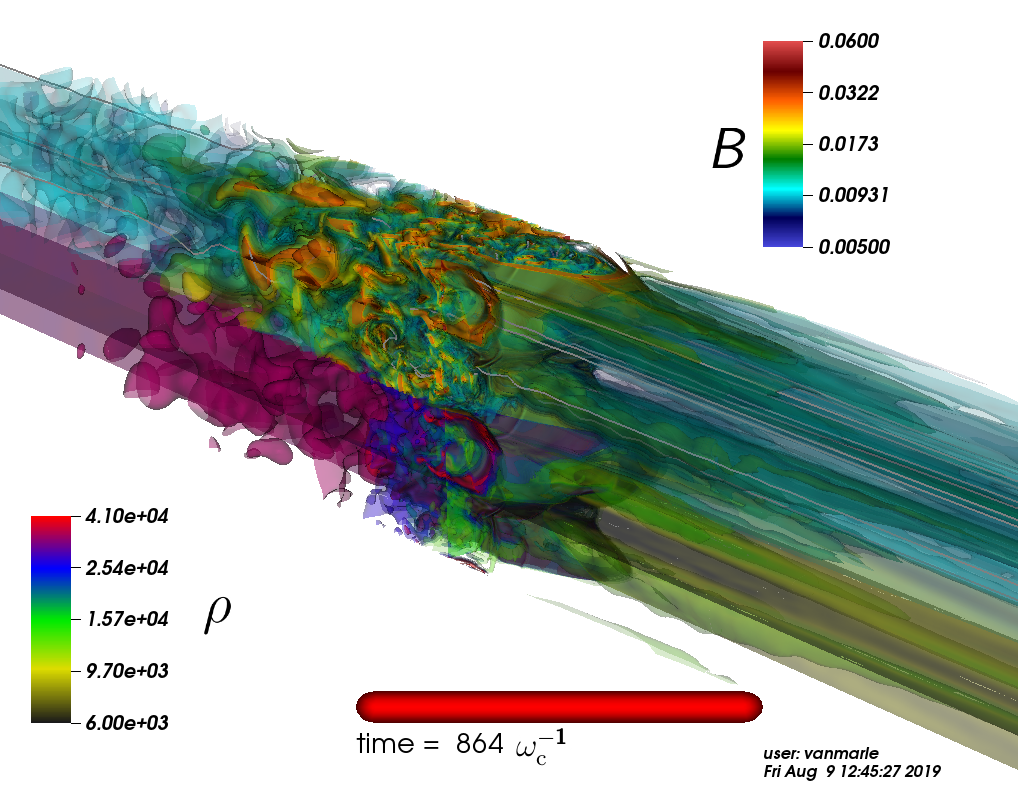}}
\caption{In this image we show an animation of the 3-D simulation of a parallel shock with $M_{\rm S}\,=\,30$. The animation shows iso-surfaces of the absolute magnetic field strength (top), the thermal gas density (bottom), and the magnetic field lines. It clearly demonstrates the development of the pre-shock instabilities and post shock turbulent zone as well as distortion of the shock surface.}
 \label{fig:animation}
\end{figure}




\bsp	
\label{lastpage}
\end{document}